\documentstyle[aps,floats,epsfig]{revtex}

\begin{document}

\twocolumn[\hsize\textwidth\columnwidth\hsize\csname 
@twocolumnfalse\endcsname

\draft
\title{Configuration of separability and tests for multipartite entanglement in Bell-type experiments}
\author{Koji Nagata, Masato Koashi, and Nobuyuki Imoto}
\address{
CREST Research Team for Interacting Carrier Electronics,
School of Advanced Sciences,\\ The Graduate University for
Advanced Studies (SOKENDAI), Hayama, Kanagawa, 240-0193, Japan
}
\date{\today}
\maketitle

\begin{abstract}
We derive tight quadratic inequalities for 
all kinds of hybrid separable-inseparable $n$-particle density 
operators on an arbitrary dimensional space. 
This methodology enables us to truly derive a
tight quadratic inequality as tests for full $n$-partite entanglement
in various Bell-type correlation experiments on the systems that may not
 be identified as a collection of qubits, e.g., those involving
photons measured by incomplete detectors.
It is also proved that when the two measured observables are assumed to precisely anti-commute, a stronger quadratic inequality can be used as a witness 
of full $n$-partite entanglement.
\pacs{PACS numbers: 03.65.Ud}
\end{abstract}
]
\narrowtext


Since 1980's, 
it has been a problem 
how to confirm multipartite entanglement experimentally.
Recently, we have been given precious experimental data by efforts of 
experimentalists\cite{bib:GHZexp,bib:pan}.
Proper analysis of these experimental data then becomes necessary, and 
as a result of such analysis\cite{bib:nagata}, the experimental data 
obtained by Pan and co-workers\cite{bib:pan} 
confirms the existence of genuinely three-particle 
entanglement under the assumption that proper observables are measured 
in the experiment.
However, it was discussed \cite{bib:Seevinck} that for other experimental data there is a loophole problem in confirming three-particle entanglement, and the 
loophole problem remains unresolved.
This means that there have not been enough 
discussions about what kind of data 
are needed for confirming multipartite entanglement.

There have been many researches 
on the problem that provide inequalities for functions 
of experimental
correlations\cite{bib:nagata,bib:Seevinck,bib:Svetlichny,bib:Gisin3,bib:collins,bib:See2,bib:Uffink,bib:Werner2}.
Among them, 
assuming $k$-partite split of the system\cite{bib:Dur2} 
without assuming a specific partition,
Werner and Wolf derived an upper bound $2^{(n-k)/2}$ for 
expectation values of $n$-particle Bell-Mermin 
operators ${\cal B},{\cal B}'$\cite{bib:Werner2,bib:Mermin} under
 the assumption that suitable partial transposes of the density operator
are positive\cite{bib:Werner2}. The inequality derived by Werner {\it et
al}. is useful because 
it tells us a number $k$ such that the given state is at most 
$k$-separable\cite{bib:Dur2,bib:com2}.

Recently, Uffink has introduced a non-linear inequality
which is aimed at giving stronger tests for full $n$-partite
entanglement than previous formulas. For qubit systems,
Uffink has derived \cite{bib:Uffink} a 
tight quadratic inequality for the states
where one qubit is not entangled with any other qubit;
namely, the states written as a convex sum over the states of
form $\rho_1\otimes\rho_{2,\ldots,n}$.

In most of real experiments, we have to deal with higher dimensional
systems rather than qubit systems. For example, when polarizations
of photons from a non-ideal source are measured by imperfect detectors,
it is difficult to claim strictly that the observed correlations
are obtained by measuring subsystems with
only two-dimensional Hilbert spaces, due to the ambiguity in the number
of photons. 
The arguments about higher dimensional systems will thus be
necessary in order to establish tests applicable to real experiments without 
making auxiliary assumptions as to the dimension of the measured space 
or as to measured observables.

In deriving a witness of full $n$-partite entanglement, 
it should be ensured that the witness rules out all hybrid separable-inseparable states except genuine fully $n$-partite entangled states.
The hybrid separable-inseparable states are depicted as follows.
Consider a partition of $n$-particle system $\{1,2,\ldots,n\}$ into $k$
nonempty and disjoint subsets 
$\alpha_1,\ldots,\alpha_k$, where 
$\sum^k_{i=1}|\alpha_i|=n$, 
to which we refer as a $k$-partite split of the system\cite{bib:Dur2}.
Let us now consider the density operators $\rho$ on 
${\cal H}=\bigotimes_{j=1}^n{\cal H}_j$, where ${\cal H}_{j}$ represents
 the Hilbert space with respect to particle $j$.
Then all hybrid separable-inseparable states with respect to
partition
$\alpha_1,\ldots,\alpha_k$ can be written as
\begin{eqnarray}
\rho=\sum_{l} p_l 
\left(\otimes^k_{i=1}\rho_l^{\alpha_i}\right), (p_l\geq 0,\sum_l p_l=1)
\label{ksep},
\end{eqnarray}
where $\rho^{\alpha_i}_l,\forall l$ are the density operators on the partial Hilbert space
$\bigotimes_{j\in\alpha_i}{\cal H}_{j}$.
States (\ref{ksep}) 
are called $k$-separable with respect to a partition 
$\alpha_1,\ldots,\alpha_k$.

In this paper, we derive the optimal upper bound of 
$\langle{\cal B}\rangle^2+\langle{\cal B}'\rangle^2$
for any partition of the systems $\alpha_1,\ldots,\alpha_k$
of an arbitrary dimensional space.
It turns out that the optimal upper bound depends only on 
two parameters $k$ and $m$, 
and not on the detailed configuration of the partition, 
where $m$ is the
number of particles which are not entangled with any other particles. 
The maximum is given by $2^{n+m-2k+1}$. 
Using this maximum, we genuinely prove that the optimal upper bound that is utilizable to confirm full $n$-partite 
entanglement ($n\geq 3$) of an arbitrary dimensional system is $2^{n-2}$.
Later, we show that if an auxiliary assumption 
as to measured observables is allowed, a stronger quadratic inequality can be used as a witness of full $n$-partite entanglement.



In what follows, we derive tight quadratic inequalities for 
hybrid separable-inseparable states with respect to partition
$\alpha_1,\ldots,\alpha_k$ of an arbitrary dimensional space.
 It is assumed that a measurement with two outcomes, 
$\pm 1$, is performed on each particle. 
Such a measurement is generally described by 
a positive-operator-valued measure (POVM), $\{F_+,F_-\}, F_++F_-={\bf
1},  F_+,F_-\geq 0$, and the corresponding  
observable is given by a Hermitian operator 
$A=F_+-F_-$, which has a spectrum in $[-1,1]$.
We assume that for each particle $j$,
 either of two such observables $A_j$ or $A'_j$ 
is chosen, 
where 
$-{\bf 1}\leq A_j,A'_j \leq {\bf 1}, \forall j$.

The Bell-Mermin operators take a simple form when 
we view on a complex plane using
a function  $f(x,y)=\frac{1}{\sqrt{2}}e^{-i\pi/4}(x+iy), 
x,y\in {\bf R}$.
Note that this function is invertible, as 
$
x=\Re f-\Im f, y=\Re f+\Im f
$.
The Bell-Mermin operators ${\cal B}_{{\bf N}_{n}}$ and
${\cal B}'_{{\bf
N}_{n}}$ are defined by\cite{bib:Werner2,bib:Mermin}
\begin{eqnarray}
f({\cal B}_{{\bf N}_{n}},{\cal B}'_{{\bf N}_{n}})=\otimes_{j=1}^n
f(A_j,A'_j),\label{MK}
\end{eqnarray}
where ${\bf N}_n=\{1,2,\ldots,n\}$.
We also define ${\cal B}_{\alpha}$ for any subset $\alpha\subset{\bf N}_n$ by
\begin{eqnarray}
f({\cal B}_{\alpha},{\cal B}'_{\alpha})=\otimes_{j\in \alpha}
f(A_j,A'_j).
\end{eqnarray}
It is easy to see, when $\alpha, \beta(\subset{\bf N}_n)$ are disjoint, that
\begin{eqnarray}
f({\cal B}_{\alpha\cup \beta},{\cal B}'_{\alpha\cup \beta})=f({\cal B}_{\alpha},{\cal B}'_{\alpha})\otimes
f({\cal B}_{\beta},
{\cal B}'_{\beta}),\label{ko}
\end{eqnarray}
which leads to
\begin{eqnarray}
{\cal B}_{\alpha\cup \beta}&=&
1/2({\cal B}_{\alpha}{\cal B}'_{\beta}
+{\cal B}'_{\alpha}{\cal B}_{\beta})
+
1/2({\cal B}_{\alpha}{\cal B}_{\beta}-
{\cal B}'_{\alpha}{\cal B}'_{\beta}),\nonumber\\
{\cal B}'_{\alpha\cup \beta}&=&
1/2({\cal B}_{\alpha}{\cal B}'_{\beta}
+{\cal B}'_{\alpha}{\cal B}_{\beta})
-
1/2({\cal B}_{\alpha}{\cal B}_{\beta}-
{\cal B}'_{\alpha}{\cal B}'_{\beta}).\label{GB}
\end{eqnarray}
First, we prove that the 
following inequality proposed by Uffink for qubit systems are also valid for an arbitrary dimensional system:
\begin{eqnarray}
\langle{\cal B}_{\alpha}\rangle^2
+
\langle{\cal B}'_{\alpha}\rangle^2
\leq 2^{|\alpha|-1}, (|\alpha|\geq 2).\label{Uffink}
\end{eqnarray}
In order to see this, we use the following lemma.


{\it Lemma: Let 
$-{\bf 1}\leq X_i,X'_i\leq{\bf 1}$ be Hermitian operators 
($i=1,2$), and define $Y,Y'$ as follows,
\begin{eqnarray}
f(Y,Y')=f(X_1,X'_1)\otimes f(X_2,X'_2).
\end{eqnarray}
Then 
\begin{eqnarray}
\langle Y\rangle^2+\langle Y'\rangle^2\leq 2.
\end{eqnarray}
}
The lemma is proven in the following way: 
Note that
\begin{eqnarray}
Y&=&(1/2)\{X_1(X_2+X'_2)+X'_1(X_2-X'_2)\},\nonumber\\
Y'&=&(1/2)\{X'_1(X'_2+X_2)+X_1(X'_2-X_2)\},
\end{eqnarray}
and let $B_{\theta}$ be $Y\cos\theta+Y'\sin\theta$. Let us derive 
the maximum value of ${\rm tr}[\rho B_{\theta}]$.
Note that
${\rm tr}[\rho B_{\theta}]$ is a 
linear function of each $X_i$ or $X'_i$, 
keeping $\rho$ fixed.
Therefore we may consider only the set of extremal points in the 
convex set of Hermitian operators with
$-{\bf 1}\leq X\leq{\bf 1}$.
Hence we may assume $X_i^2={X'_i}^2={\bf 1}(i=1,2)$.
We thus have
\begin{eqnarray}
Y^2={Y'}^2&=&{\bf 1}-(1/4)[X_1,X'_1]\otimes[X_2,X'_2]\nonumber\\
&=&{\bf 1}+A^-_1\otimes A^-_2,
\end{eqnarray}
where $A_i^-=(i/2)[X_i,X'_i]$ ($A_i^-$ are Hermitian operators) and
\begin{eqnarray}
\{Y,Y'\}=(1/2)\{X_1,X'_1\}\otimes\{X_2,X'_2\}
=2A^+_1\otimes A^+_2,
\end{eqnarray}
where $A_i^+=(1/2)\{X_i,X'_i\}$ ($A_i^+$ are Hermitian operators).
Then we obtain
\begin{eqnarray}
B^2_{\theta}={\bf 1}+A^-_1\otimes A^-_2+\sin 2\theta A^+_1\otimes A^+_2.\label{cal}
\end{eqnarray}
This implies 
\begin{eqnarray}
{\rm tr}[\rho B^2_{\theta}]\leq 
&&\max\{1+\Vert A^+_1\otimes A^+_2+ A^-_1\otimes A^-_2\Vert,\nonumber\\
&&\quad1+\Vert A^+_1\otimes A^+_2- A^-_1\otimes A^-_2\Vert\},
\end{eqnarray}
where $\Vert \cdot \Vert$ means the operator norm.
Note that
\begin{eqnarray}
&&A^+_1\otimes A^+_2\pm A^-_1\otimes A^-_2\nonumber\\
&=&(1/2)\{(A^+_1+iA^-_1)\otimes(A^+_2\mp iA^-_2)\nonumber\\
&&\quad+
(A^+_1-iA^-_1)\otimes(A^+_2\pm iA^-_2)\},
\end{eqnarray}
and
\begin{eqnarray}
&&A^+_1\otimes A^+_2+ A^-_1\otimes A^-_2\nonumber\\
&=&(1/2)(X'_1X_1 \otimes X_2X'_2+X_1X'_1 \otimes X'_2X_2).
\end{eqnarray}
According to relationships such as
$
\Vert X'_1X_1\Vert=\Vert(X_1X'_1)(X'_1X_1)\Vert^{1/2}=\Vert{\bf 1}\Vert^{1/2}=1
$,
we get
\begin{eqnarray}
\Vert A^+_1\otimes A^+_2+ A^-_1\otimes A^-_2\Vert\leq 1.
\end{eqnarray}
Similarly, we also get
\begin{eqnarray}
&&\Vert A^+_1\otimes A^+_2- A^-_1\otimes A^-_2\Vert\nonumber\\
&=&(1/2)\Vert X'_1X_1 \otimes X'_2X_2+X_1X'_1 \otimes X_2X'_2\Vert
\leq 1.
\end{eqnarray}
Therefore we 
have $|{\rm tr}[\rho B_{\theta}]|^2\leq {\rm tr}[\rho B^2_{\theta}]\leq 2$ by 
the variance inequality.
Now by taking 
\begin{eqnarray}
\cos \theta=\frac{\langle Y\rangle}
{\sqrt{\langle Y\rangle^2+\langle Y'\rangle^2}},\ 
\sin \theta=\frac{\langle Y'\rangle}
{\sqrt{\langle Y\rangle^2+\langle Y'\rangle^2}},
\end{eqnarray}
we obtain
$
\langle Y\rangle^2+\langle Y'\rangle^2\leq 2$, Q.E.D.


Let us consider a set $\alpha\subset{\bf N}_n$, where $|\alpha|\geq 2$.
Let $\gamma$ be $\alpha\backslash\{j\}$, where $j\in \alpha$.
Then, from Eq.~(\ref{ko}), we have
\begin{eqnarray}
f({\cal B}_{\alpha},{\cal B}'_{\alpha})
=
f({\cal B}_{\gamma},
{\cal B}'_{\gamma})\otimes
f(A_j,A'_j).
\end{eqnarray}
It has been known that the maximum of $\langle {\cal B}_{\gamma} \rangle$ is 
$2^{(|\gamma|-1)/2}$\cite{bib:Werner2}.
Noting that $f(cx,cy)=cf(x,y), c\in{\bf R}$ and 
$-{\bf 1}\leq 2^{-(|\gamma|-1)/2}{\cal B}_{\gamma} \leq{\bf 1}$,
according to the lemma by taking $X_1=2^{-(|\gamma|-1)/2}{\cal B}_{\gamma}, 
X'_1=2^{-(|\gamma|-1)/2}{\cal B}'_{\gamma}, X_2=A_j$, and $X'_2=A'_j$,
we obtain the quadratic inequality
\begin{eqnarray}
\langle{\cal B}_{\alpha}\rangle^2
+
\langle{\cal B}'_{\alpha}\rangle^2
\leq 2^{|\alpha|-1},\ (|\alpha|\geq 2),
\end{eqnarray}
where we used $|\alpha|=|\gamma|+1$.

Next, we calculate an upper bound of $\langle{\cal B}_{{\bf N}_n}\rangle^2
+
\langle{\cal B}'_{{\bf N}_n}\rangle^2$ for states of the form
$\otimes^k_{i=1}\rho^{\alpha_i}$. From Eq.~(\ref{GB}), we have
\begin{eqnarray}
&&\langle{\cal B}_{\alpha\cup\beta}\rangle^2
+
\langle{\cal B}'_{\alpha\cup \beta}\rangle^2\nonumber\\
&&=(1/2)\big(\langle{\cal B}_{\alpha}
{\cal B}'_{\beta}+
{\cal B}'_{\alpha}
{\cal B}_{\beta}\rangle^2
+
\langle{\cal B}_{\alpha}
{\cal B}_{\beta}-{\cal B}'_{\alpha}
{\cal B}'_{\beta}\rangle^2\big).\label{KN}
\end{eqnarray}
Using Eq.~(\ref{KN}), we obtain
\begin{eqnarray}
&&\langle{\cal B}_{{\bf N}_n}\rangle^2
+
\langle{\cal B}'_{{\bf N}_n}\rangle^2\nonumber\\
&&=(1/2)\big(\langle{\cal B}_{\alpha_1}
{\cal B}'_{{\bf N}_{n}\backslash{\alpha_1}}+
{\cal B}'_{\alpha_1}
{\cal B}_{{\bf N}_{n}\backslash{\alpha_1}}\rangle^2
\nonumber\\
&& \quad +
\langle{\cal B}_{\alpha_1}
{\cal B}_{{\bf N}_{n}\backslash{\alpha_1}}-{\cal B}'_{\alpha_1}
{\cal B}'_{{\bf N}_{n}\backslash\alpha_1}\rangle^2\big)\nonumber\\
&&=(1/2)\big\{\big(\langle{\cal B}_{\alpha_1}\rangle
\langle{\cal B}'_{{\bf N}_{n}\backslash{\alpha_1}}\rangle+
\langle{\cal B}'_{\alpha_1}\rangle
\langle{\cal B}_{{\bf N}_{n}\backslash{\alpha_1}}\rangle\big)^2
\nonumber\\
&& \quad +
\big(\langle{\cal B}_{\alpha_1}\rangle
\langle{\cal B}_{{\bf N}_{n}\backslash{\alpha_1}}\rangle
-\langle{\cal B}'_{\alpha_1}\rangle
\langle{\cal B}'_{{\bf N}_{n}\backslash\alpha_1}\rangle\big)^2\big\}
\nonumber\\
&&=(1/2)\big(\langle{\cal B}_{\alpha_1}\rangle^2+
\langle{\cal B}'_{\alpha_1}\rangle^2\big)
\big(\langle{\cal B}_{{\bf N}_{n}\backslash{\alpha_1}}\rangle^2+
\langle{\cal B}'_{{\bf N}_{n}\backslash{\alpha_1}}\rangle^2\big).
\end{eqnarray}
Repeating this, we obtain
\begin{eqnarray}
\langle{\cal B}_{{\bf N}_n}\rangle^2
+
\langle{\cal B}'_{{\bf N}_n}\rangle^2
=(1/2)^{k-1}\prod_{i=1}^k(\langle{\cal B}_{\alpha_i}\rangle^2
+\langle{\cal B}'_{\alpha_i}\rangle^2).
\end{eqnarray}
Without loss of generality, we assume that $|\alpha_i|=1$ 
for $1\le i \le m$ 
and $|\alpha_i|\geq 2$ for $m+1\le i \le k$. Applying $\langle{\cal
B}_{\alpha_i}\rangle^2 +\langle{\cal B}'_{\alpha_i}\rangle^2\leq 2$ for
$|\alpha_i|=1$ and  Eq.~(\ref{Uffink}),
we obtain
\begin{eqnarray}
&&\langle{\cal B}_{{\bf N}_n}\rangle^2
+
\langle{\cal B}'_{{\bf N}_n}\rangle^2\nonumber\\
&&\leq \prod_{i=m+1}^k 2^{(|\alpha_{i}|-1)}(1/2)^{k-m-1}=2^{n+m-2k+1},
\end{eqnarray}
where we used $\sum_{i=m+1}^{k}(|\alpha_i|-1)=(n-m)-(k-m)$.
We then conclude\cite{bib:com1} that, for any state $\rho$ 
that is $k$-separable with respect to $\alpha_1,\ldots,\alpha_k$,
\begin{eqnarray}
({\rm tr}[\rho {\cal B}_{{\bf N}_n}])^2+
({\rm tr}[\rho {\cal B}'_{{\bf N}_n}])^2
\leq 2^{n+m-2k+1}.\label{final}
\end{eqnarray}
The maximum depends only on two parameters $k$ and $m$ 
but not on the detailed configuration of the partition. 
Cleary the bound (\ref{final}) is optimal.

The inequality for testing full $n$-partite entanglement for $n\geq 3$ is obtained by maximizing the right-hand side of (\ref{final}) under 
the condition $k\geq 2$.
Noting that $m\leq k-1$ when $k<n$, we obtain
\begin{eqnarray}
\langle{\cal B}_{{\bf N}_n}\rangle^2
+
\langle{\cal B}'_{{\bf N}_n}\rangle^2
\leq 2^{n-2},\label{Uffink2}
\end{eqnarray}
Violations of the inequality (\ref{Uffink2}) imply
full $n$-partite entanglement.

For multiqubit systems, Uffink considered the case 
of partitions of the form $\{1\},\{2,\ldots,n\}$, 
and has presented the quadratic inequality (\ref{Uffink2}) for testing 
whether $n$-particle states are fully entangled\cite{bib:Uffink}. 
In what we should pay attention to,
 we have to check that for all hybrid separable-inseparable states except genuine fully entangled states, the optimal
upper bounds 
are smaller than or equal to $2^{n-2}$,
 in order to ensure that the relation
(\ref{Uffink2}) can be used as tests for full $n$-partite entanglement.
In this point, we genuinely proved that the violations
of the relation  (\ref{Uffink2}) are a sufficient for confirming
fully $n$-partite entangled states. 
We have also proven that the relation (\ref{Uffink2}) 
can be derived not only for multiqubit systems 
but also for higher dimensional systems.

The inequality (\ref{final}) also implies
\begin{eqnarray}
|{\rm tr}[\rho {\cal B}_{{\bf N}_n}]|\leq 2^{(n+m-2k+1)/2}.
\end{eqnarray}
It is known that $|{\rm tr}[\rho {\cal B}_{{\bf N}_n}]|\leq 1$ when 
the system is fully separable\cite{bib:Werner2}.
Hence we obtain an upper bound\cite{bib:com10}
\begin{eqnarray}
|{\rm tr}[\rho {\cal B}_{{\bf N}_n}]|
\leq 
\left\{
\begin{array}{cl}
\displaystyle
2^{(n+m-2k+1)/2}
&\quad k<n\\
\\
\displaystyle
1
&\quad k=n.
\end{array} \right.\label{KN2}
\end{eqnarray}
According to Eq.~(\ref{GB}), the equality of the relation 
(\ref{KN2}) holds when 
$\langle {\cal B}_{\alpha_i}\rangle=
\langle {\cal B}'_{\alpha_i}\rangle=1$ for $1\le i\le m$, 
$\langle{\cal B}_{\alpha_i}\rangle
=\langle{\cal B}'_{\alpha_i}\rangle=2^{(|\alpha_{i}|-2)/2}$ 
for $m+1\le i\le k-1$, and
$\langle{\cal B}_{\alpha_k}\rangle
=2^{(|\alpha_{k}|-1)/2}$.
We can find a state and Hermitian 
operators $-{\bf 1}\leq A_j,A'_j\leq {\bf 1}$ that satisfy the above relations\cite{bib:com8}.
Hence the bound (\ref{KN2}) is optimal.

For partitions of 
the form $\{1\},\{2\},\ldots,\{m\},\{m+1,\ldots,n\}(m\leq n-1)$, 
the relation (\ref{KN2}) leads to 
the result of Gisin and Bechmann-Pasquinucci\cite{bib:Gisin3}, 
i.e., the
bound 
$|\langle{\cal B}_{{\bf N}_n}\rangle|\leq 2^{(n-m-1)/2}$.
Noting that $m\leq k-1$ when $k<n$,
the relation (\ref{KN2}) leads to 
the result of Werner and Wolf\cite{bib:Werner2}, i.e., 
$|\langle{\cal B}_{{\bf N}_n}\rangle|\leq 2^{(n-k)/2}$ 
by taking the maximum over $m$ with fixed $k$.
Collins {\it et al}. considered 
the cases for 
partitions of the form 
$\{1\},\{2\},\{3,4\}$ or $\{1,2\},\{3,4\}$ or $\{1\},\{2,3,4\}$
and presented the bounds as $\sqrt{2},\sqrt{2},2$, 
respectively\cite{bib:collins}.
These bounds are also derived from the relation (\ref{KN2}).

So far, we derived the threshold value (i.e., $2^{n-2}$) of 
$\langle{\cal B}_{{\bf N}_n}\rangle^2
+
\langle{\cal B}'_{{\bf N}_n}\rangle^2$ for use as a full $n$-partite 
entanglement witness 
over all observables satisfying 
$-{\bf 1}\leq A_j,A'_j\leq {\bf 1}$.
Now, let us consider an additional assumption 
that the two measured observables anti-commute,
i.e., $\{A_j,A'_j\}={\bf
0}\forall j$.  
 This assumption is
approximately fulfilled within the accuracy of the measurement
apparatus in the common experimental situations, e.g.,
when we measure Pauli operators $\sigma_x$ and $\sigma_y$ for
each particle. 
With this assumption, the threshold
value of 
$\langle{\cal B}_{{\bf N}_n}\rangle^2 + \langle{\cal B}'_{{\bf N}_n}\rangle^2$
becomes as small as $2^{n-3}$ as is shown below.
This implies that we can use a stronger quadratic
inequality as tests for full $n$-partite entanglement in this case.

Suppose that 
$\{A_j,A'_j\}={\bf 0}$ and 
$-{\bf 1}\leq A_j,A'_j \leq {\bf 1}, \forall j$.
Let us take
$
{A_j}_{\theta}=A_j\cos \theta +A'_j\sin \theta,
$
and derive the maximum of the values ${\rm tr}[\rho{A_j}_{\theta}]$.
Since we are only interested in the maximum,
we may assume ${A_j}^2={A'_j}^2={\bf 1}$.
Then we get
$
{A_j}^2_{\theta}={\bf 1}+(1/2)\{A_j,A'_j\}\sin 2\theta={\bf 1}.
$
The variance inequality leads to
$
|{\rm tr}[\rho {A_j}_{\theta}]|^2\leq {\rm tr}[\rho {A_j}^2_{\theta}]=1.
$
Now take
$
\cos \theta=\langle A_j\rangle/
\sqrt{\langle A_j\rangle^2+\langle A'_j\rangle^2},\ 
\sin \theta=\langle A'_j\rangle/
\sqrt{\langle A_j\rangle^2+\langle A'_j\rangle^2},
$
then we get
$
\langle A_j\rangle^2+\langle A'_j\rangle^2\leq 1\forall j.
$
This means that the relation (\ref{Uffink}) holds even for $|\alpha|=1$.
Hence we obtain
\begin{eqnarray}
\langle{\cal B}_{\alpha}\rangle^2
+
\langle{\cal B}'_{\alpha}\rangle^2
\leq 2^{|\alpha|-1}, (|\alpha|\geq 1).\label{KN3}
\end{eqnarray}
Similar to the argument that derives (\ref{final}), applying 
the relation (\ref{KN3}), we conclude 
\begin{eqnarray}
({\rm tr}[\rho {\cal B}_{{\bf N}_n}])^2
+
({\rm tr}[\rho {\cal B}'_{{\bf N}_n}])^2\leq 2^{n-2k+1}.\label{quad}
\end{eqnarray}

The inequality for testing full $n$-partite entanglement is obtained by maximizing the right-hand side of (\ref{quad}) under 
the condition $k\geq 2$.
We obtain\cite{bib:com11}
\begin{eqnarray}
\langle{\cal B}_{{\bf N}_n}\rangle^2
+
\langle{\cal B}'_{{\bf N}_n}\rangle^2\leq 2^{n-3}.\label{NSIM}
\end{eqnarray}

We give an example that the relation (\ref{NSIM}) is stronger than (\ref{Uffink2}) as a witness of full $n$-partite entanglement for multiqubit systems.
We assume that
$A_j=\vec{a_j}\cdot\vec{\sigma},A'_j=\vec{a'_j}\cdot\vec{\sigma}$, where
$\vec{a_j}$ and $\vec{a'_j}$ are normalized vectors in ${\bf R}^3$ and 
$\vec{\sigma}$ is the vector of Pauli matrices, i.e., 
$\vec{\sigma}=(\sigma_x,\sigma_y,\sigma_z)$.
The condition $\{A_j,A'_j\}={\bf 0}$ leads to
$\vec{a_j}\cdot\vec{a'_j}=0$.
Let us consider the following multiqubit states\cite{bib:Dur2}:
\begin{eqnarray}
\rho=x|\Phi_n\rangle\langle\Phi_n|+\frac{1-x}{2^n}I,\label{Wstate}
\end{eqnarray}
where $I$ is the identity operator for the $2^n$-dimensional space and
$|\Phi_n\rangle$ is an $n$-qubit GHZ state\cite{bib:GHZ}, i.e.,
\begin{eqnarray}
|\Phi_n\rangle=\frac{1}{\sqrt{2}}(|+_1,+_2, \ldots ,+_n\rangle+
|-_1,-_2,\ldots,-_n\rangle).
\end{eqnarray}
It is easy to see that the maximum of 
$\langle{\cal B}_{{\bf N}_n}\rangle^2
+
\langle{\cal B}'_{{\bf N}_n}\rangle^2$ is $2^{n-1} x^2$ with 
$\vec{a_j}\cdot\vec{a'_j}=0\forall j$ (See \cite{bib:Werner3}).
Hence,
assuming that $x$ is in the range of 
\begin{eqnarray}
\frac{1}{2}<x\leq \frac{1}{\sqrt{2}},
\end{eqnarray}
we can confirm full
$n$-partite entanglement from (\ref{NSIM}), which cannot be
confirmed by (\ref{Uffink2}).
Hence if the measurement setups are precisely chosen as 
$\{A_j,A'_j\}={\bf 0}\forall j$, 
then one can use a stronger 
inequality as tests for full $n$-partite entanglement
in comparison with the relation (\ref{Uffink2}).

In real experimental situations, we cannot claim that $\{A_j,A'_j\}={\bf 0}$ with arbitrary precision.
The relevance of the bounds claiming full $n$-partite entanglement
assuming that $|\langle\{A_j,A'_j\}\rangle|\leq\epsilon$, 
where $\epsilon$ means experimental errors, would be worth further investigations.


In summary, we have derived the quadratic inequality that is utilizable
to test full $n$-partite entanglement not only for qubit systems
but also for higher dimensional systems. This helps the analysis of
experimental data in realistic situations. 
We have also shown that when
the two measured observables are assumed to precisely anti-commute, we 
can use a stronger quadratic  
inequality as a witness of full $n$-partite
entanglement in  correlation experiments.

K.N. thanks Masanao Ozawa for helpful discussions.





\begin{thebibliography}{99}
\bibitem{bib:com2}

We can estimate the minimum number of entangled particles as 
$
n/k,
(n/k\in{\bf N})$ or $
\left[n/k\right]+1,
(n/k\not\in{\bf N}) 
$, 
when we are 
given the maximum positive integer $k$ 
which is allowed by experimental data.
\bibitem{bib:com1}
As have been discussed in Ref.~\cite{bib:Uffink}, 
as to any Hermitian operator ${\cal O}$, 
the expression
$({\rm tr}[\rho{\cal O}])^2$
is a convex function of $\rho$, and it is 
sufficient to consider only the states $\otimes_{i=1}^k\rho^{\alpha_i}$  
to obtain the optimal upper bound for states (\ref{ksep}).
\bibitem{bib:com10}
One can rewrite the term in the exponent as 
$n+m-2k+1=n-m-2(k-m)+1=n'-2k'+1$, where $n'$ and $k'$ are respectively the number of particles and the number of partitions that are left when
disentangled $m$ particles are left aside.
\bibitem{bib:com11}
What we have shown is that the relations (\ref{Uffink2}) and (\ref{NSIM}) 
are entanglement witnesses.
It is still open whether the relation (\ref{Uffink2}) can 
be used to rule out hidden variable models.
\end{thebibliography}
\end{document}